\documentclass[fp,twocolumn]{jpsj3}

\usepackage{txfonts}
\usepackage{graphicx}
\usepackage{bm}
\usepackage{xcolor}
\usepackage{ulem}

\begin{document}

\title{
Unusual upper critical field in UTe$_2$ revealed by magnetotransport measurements up to 42 T \\
}

\author{Macha M\'{e}plan$^1$, Ilya Sheikin$^2$, Pierre Pugnat$^2$, Romain Barbier$^2$, Fran\c{c}ois Debray$^2$, C\'{e}dric 
 Grandcl\'{e}ment$^2$, Steffen Kr\"{a}mer$^2$, Yuriy Krupko$^2$, Fr\'{e}d\'{e}ric Molini\'{e}$^3$, K\'{e}vin Paillot$^2$, Robert Pankow$^2$, Rolf Pfister$^2$, Luc Ronayette$^2$, Benjamin Vincent$^2$, Charles Simon$^2$, Midori Amano Patino$^{1,4}$, G\'{e}rard Lapertot$^1$, Georg Knebel$^1$\thanks{georg.knebel@cea.fr}, Dai Aoki$^5$\thanks{dai.aoki.c2@tohoku.ac.jp}}

\inst{$^1$Univ. Grenoble Alpes, CEA, Grenoble INP, IRIG, PHELIQS, F-38000 Grenoble, France \\
$^2$Laboratoire National des Champs Magn\'{e}tiques Intenses, LNCMI-CNRS, EMFL, Univ. Grenoble Alpes, Univ. Toulouse, INSA-T, 38042 Grenoble, France \\
$^3$IRFU, CEA Paris-Saclay,  91191 Gif-sur-Yvette, France \\
$^4$Univ.~Grenoble Alpes, CNRS, Grenoble-INP, Institut N\'eel, 38000 Grenoble, France\\
$^5$Institute for Materials Research, Tohoku University, Oarai, Ibaraki, 311-1313, Japan} 

\abst{
The heavy-fermion superconductor UTe$_2$ is unique in that, at ambient pressure, it exhibits three distinct superconducting phases, two of which are induced by magnetic field. When the field is applied along the crystallographic $\bm{b}$ axis in the orthorhombic structure, the field-induced phase SC2 develops above approximately 20 T and persists up to the metamagnetic transition at $H_m \sim 34$~T. When the magnetic field is tilted towards the $\bm{c}$ axis, another superconducting phase, SC3, emerges at very high fields above about 40 T over a certain angular range. The origin of this exotic phase remains under debate. One of the key open questions regarding the origin of SC3 is whether it is confined to the spin-polarized state above $H_m$, or whether it already develops at lower fields. Here, we report magnetoresistance measurements performed on a high-quality single crystal of UTe$_2$ in static magnetic fields up to 42~T applied in the ($\bm{bc}$) plane at temperatures down to 0.35~K. At this temperature, we find that the SC3 phase first appears at an angle of 20 deg from the $\bm{b}$ axis. At larger angles, the onset of the SC3 phase, defined by a maximum in resistivity, occurs below $H_m$. However, zero resistivity is reached only above $H_m$ throughout the entire angular range investigated. These results are summarized in the resulting field-angle phase diagram. Furthermore, we find that at 21 deg the SC3 phase is rapidly suppressed with increasing temperature, whereas at 24 deg it becomes considerably more robust and persists up to about 1~K. Finally, we observe Shubnikov–de Haas (SdH) oscillations in the vicinity of the $\bm{c}$ axis. The observed oscillation frequencies are in good agreement with our previous results. The field dependence of the strongest SdH frequency and of the effective mass is discussed.
}


\maketitle

\section{Introduction}

Since the discovery of unusual, presumably spin-triplet, superconductivity in the heavy-fermion compound UTe$_2$ slightly more than seven years ago \cite{Ran2019a, Aoki2019}, this material has attracted intense experimental and theoretical interest. UTe$_2$ crystallizes in a body-centered orthorhombic structure (space group: No. 71, $Immm$, $D_{2h}^{25}$) in which the uranium atoms form two-leg ladders along the $\bm{a}$ axis. The ground state of UTe$_2$ is paramagnetic, with a strongly enhanced Sommerfeld coefficient $\gamma \simeq 120$~mJ\,K$^{-2}$\,mol$^{-1}$. The superconducting transition temperature $T_c$ varies from about 1.6~K in lower-quality crystals to approximately 2.1~K in the highest-quality samples currently available \cite{Sakai2022, Rosa2022, Aoki2024a, Wu2024PNAS, Hayes2025}.

One of the most remarkable features of UTe$_2$ is the emergence of multiple superconducting phases under pressure \cite{Braithwaite2019, Aoki2020a, Thomas2020a, Kinjo2023a, Wu2025c} and magnetic field \cite{Ran2019, Rosuel2023, Lewin2023, Helm2024, Wu2025}. In the latter case, when the magnetic field is applied along the magnetization hard $\bm{b}$ axis, the upper critical field $H_{c2}$ exhibits reentrant behavior \cite{Ran2019, Knebel2019}. At low temperatures, the zero-field superconducting phase SC1 is suppressed at about 20~T, giving rise to another superconducting phase SC2 that persists up to the first-order metamagnetic transition at $H_m \sim 34$~T.  While the low-field SC1 phase is widely believed to possess a spin-triplet order parameter, the symmetry of the SC2 phase remains an open question.\cite{Rosuel2023, Kinjo2023, Lewin2023, Zhang2025} Furthermore, a mixed superconducting phase (SC1+SC2) has been reported in the field range separating the low-field SC1 and high-field SC2 phases.\cite{Sakai2023, Tokiwa2023, Valiska2026} The metamagnetic transition field $H_m$ is strongly angle dependent and increases rapidly when the magnetic field is rotated toward either the $\bm{a}$ or the $\bm{c}$ axis. In the latter case, somewhat surprisingly, yet another field-induced superconducting phase SC3 emerges above $H_m$ in high-quality samples within a narrow angular range $20~\rm{deg} \lesssim \theta \lesssim 45$~deg, where $\theta$ denotes the angle of the magnetic field measured from $\bm{b}$ towards the $\bm{c}$ axis.\cite{Ran2019a, Knafo2020, Lewin2023, Helm2024, Wu2025, Thebault2026} The SC3 phase is extremely robust against magnetic field and persists up to fields exceeding 70~T. This phase, originally detected in numerous contacted and contactless resistivity measurements performed in both pulsed and static magnetic fields, has recently also been observed in magnetocaloric-effect \cite{Schoenemann2024} and ultrasound measurements \cite{Marquardt2025}, suggesting the bulk nature of this superconducting state.

The origin of the high-field-induced superconducting phase SC3 remains under debate. Two main scenarios have been proposed. One invokes a field-compensation mechanism analogous to the Jaccarino--Peter effect.\cite{Helm2024} Several experimental observations support this interpretation. First, most reported measurements indicate that the high-field superconducting phase exists only in the spin-polarized state above $H_m$. Second, a strong suppression of the normal-state Hall effect has been observed around $\theta \approx 30$~deg, suggesting partial compensation of the applied magnetic field by an internal exchange field.\cite{Helm2024} Third, the SC3 phase appears to be considerably less sensitive to disorder than the low-field superconductivity,\cite{Frank2024} implying that the high-field-induced superconductivity may arise from a more conventional pairing mechanism. Such a scenario would also be consistent with a field-compensation mechanism as the origin of the high-field superconductivity.

An alternative hypothesis attributes the high-field superconductivity to quantum critical, presumably magnetic, fluctuations associated with an as-yet unidentified order parameter. This is supported by a strong increase of the $T^2$-coefficient of the electrical resistivity $A$ in the angular range 30 - 40 deg in the $\bm{(bc)}$ plane.\cite{Thebault2026} Furthermore, the observation of the "halo"-like angular dependence of the SC3 phase around the $\bm{b}$ axis was argued to support a fluctuation driven superconducting pairing mechanism.\cite{Lewin2025a, Wu2025} Finally, very recently, Wu \textit{et al.} reported that the high-field superconducting phase develops at low temperatures below $H_m$ within a narrow angular range in the ($bc$) plane \cite{Wu2026}, in difference to previous pulsed fields measurements.\cite{Knafo2020, Thebault2026} In this new study, resistivity measurements were performed down to 400~mK in static magnetic fields up to 45~T. For $\theta \gtrsim 25$~deg, once the low-field superconductivity is suppressed by relatively small magnetic fields, the resistivity increases with field and reaches a maximum at fields considerably lower than $H_m$. This maximum was interpreted as marking the onset of the high-field superconducting phase SC3. Even more striking is the reported observation of zero resistivity at fields somewhat lower than $H_m$ for $25~\rm{deg} \lesssim \theta \lesssim 35~$deg. If confirmed, these results would strongly support the scenario in which quantum critical fluctuations are responsible for the high-field-induced superconductivity. 

Motivated by these findings, we performed high-field magnetotransport measurements on a high-quality sample similar to that used by Wu \textit{et al.} in Ref.~\citen{Wu2026} with particular attention paid to the precise determination of both $H_m$ and the magnetic field at which the resistivity becomes zero. While we also observe a resistivity maximum at fields lower than $H_m$ within a certain angular range in the ($\bm{bc}$) plane, our measurements show that the high-field superconducting phase, defined by the onset of zero resistivity, is always confined to the spin-polarized state above $H_m$.

Furthermore, we observed Shubnikov-de Haas (SdH) oscillations at magnetic fields above 30~T applied close to the $\bm{c}$ axis. These SdH oscillations are in good agreement with our previous results,\cite{Aoki2022osc, Aoki2023, Aoki2024} indicating quasi-2D Fermi surfaces along the $\bm{c}$ axis with a small warping. The field dependence of the observed SdH frequencies will be discussed.

\section{Experimental details}

The single crystal of UTe$_2$ studied in this work was grown using the molten-salt flux liquid-transport technique, as described in detail elsewhere~\cite{Aoki2024a}. The high quality of the sample is demonstrated by its large residual resistivity ratio (RRR) of 576, corresponding to a residual resistivity of $\rho_0 =0.57\,\mu \Omega\!\cdot\!{\rm cm}$. The zero-field superconducting transition temperature, defined by zero resistivity, is $T_c =2.05$~K. These parameters are among the best reported in the literature.

Magnetoresistance measurements were carried out using a standard four-probe AC technique with lock-in detection. An electrical current of 400~$\mu$A at a frequency of 17.65~Hz was applied along the crystallographic $\bm{a}$ axis. The sample was mounted in a $^3$He cryostat with a base temperature of 350 mK, which was equipped with a single-axis rotator directly submerged into $^3$He condensate. The $\bm{a}$ axis of the crystal was aligned with the rotation axis of the rotator, allowing rotation of the magnetic field within the (\bm{$bc$}) plane and keeping the transverse orientation. 

Magnetic fields were generated using the recently constructed hybrid magnet~\cite{Pugnat2026, Pugnat2026a} at the Laboratoire National des Champs Magn\'{e}tiques Intenses (LNCMI) in Grenoble, France. The hybrid system consists of an outer superconducting coil with an inner diameter of 1100~mm, capable of generating fields up to 8.5~T, and an inner resistive magnet. The resistive magnet comprises an outer Bitter coil, currently limited to 8.5~T, and an inner polyhelix solenoid that can presently generate fields up to 25~T. As a result, the system can currently produce magnetic fields of up to 42~T in a 34~mm room-temperature bore. The setup, which is still in the commissioning phase, is expected to reach a maximum field of 43~T in the coming years.

\begin{figure}
\includegraphics[width=0.9\linewidth]{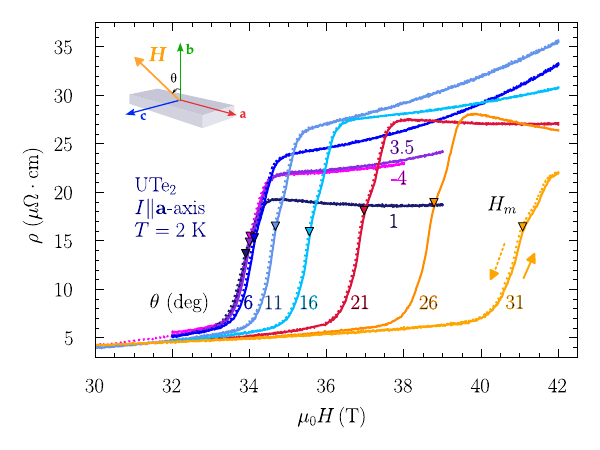}
\caption[Magnetoresistance of UTe2 at T=2 K.]{(Color online) Magnetoresistance of UTe$_2$ at $T=2~$K for different angles in the $(bc)$ plane. Solid (dashed) lines correspond to the up- (down-) field sweeps, respectively. The triangles indicate the position of $H_m$.}
\label{f1}
\end{figure}

\section{Experimental results}

\subsection{Angular dependence of the magnetoresistance}

In order to consistently determine $H_m$ at different angles, we performed up- and down-field sweeps between 30 and 42~T at $T=2$~K. At this temperature, the sample is in the normal state above 30~T for all the field angles up to at least 31 deg. The result for angles from -4 deg to 31 deg in the ($bc$) plane is shown in Fig.~\ref{f1}. Curves at -4 and 3.5 deg almost fall on top of each other demonstrating an excellent alignment of the $\bm{a}$ axis along the rotation axis. For $\bm{H} \parallel \bm{b}$, the metamagnetic transition occurs at  $\mu_0 H_m \approx 34$~T. At this orientation, the transition width is about 1.2~T; it increases significantly with increasing angle. For all the angles, $H_m$ is defined as the middle of the jump in magnetoresistance. The transition is larger than previously determined in samples grown by the chemical vapor transport (CVT). On the other hand, the hysteresis between up- and down-field sweeps, about 0.05~T (at a typical sweep rate of 150~Gauss/s), is much smaller compared to the first measurements on CVT samples, where it is $\approx 0.25$~T. The jump in magnetoresistance at the metamagnetic transition is the smallest for the field applied along the $\bm{b}$ axis. It increases up to an angle of 11~deg, but remains constant at larger angles. This behavior is different from the angular dependence of the magnetoresistance below $H_m$, where $\rho (\theta)$ shows a maximum for $H \parallel b$ both in the experiment and theoretical calculations.\cite{Kimata2026, Ishizuka2026} 
\begin{figure}
    \includegraphics[width=0.9\linewidth]{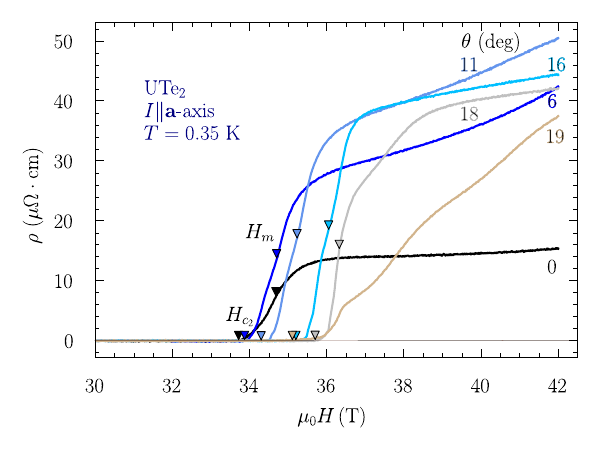}
    \includegraphics[width=0.9\linewidth]{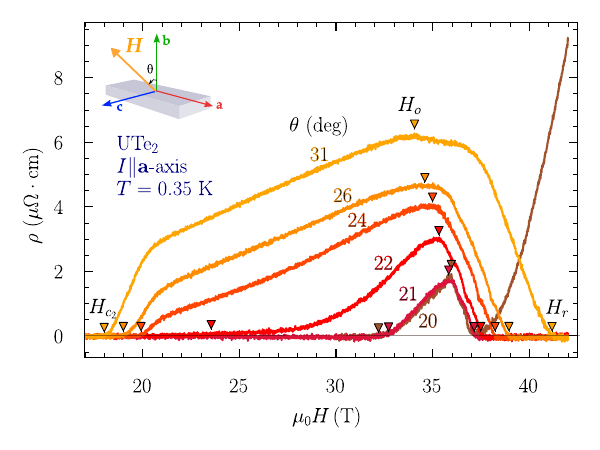}
    \includegraphics[width=0.9\linewidth]{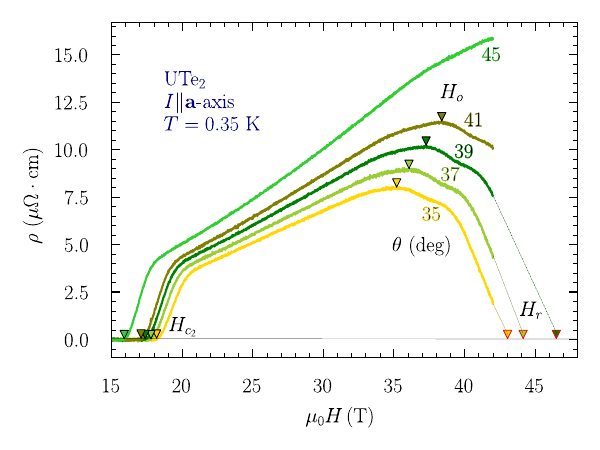}
    \caption[Magnetoresistance of UTe2 at T=0.35 K.]{(Color online) Magnetoresistance of UTe$_2$ at $T=0.35~$K is shown for different angles from $\mathbf{b}$ towards the $\mathbf{c}$ axis: $0 - 19~$deg in the upper panel, $20 - 31~$deg in the middle panel, and $35 - 45$~deg in the lower panel. Triangles mark the limit of zero resistivity of the SC2 superconducting phase and the metamagnetic transition at $H_m$ (upper panel), the upper limit of zero resistivity, and the maximum of $\rho (H)$ at $H_{o}$ (middle panel), and the upper limit of the low field superconducting phase SC1, the maximum of the magnetoresistance, and the extrapolated values of zero resistivity of the SC3 superconducting phase (lower panel).}
    \label{f2}
\end{figure}
In Fig.~\ref{f2} we present the magnetoresistance at $T=0.35$~K at different angles from $\mathbf{b}$ to the $\mathbf{c}$ axis. The upper panel shows the data for angles up to 19 deg. Superconductivity, defined as $\rho = 0$, is observed for all angles up to the onset of the metamagnetic transition. Over this angular range, the field-reinforced superconducting phase SC2 is not destroyed by the magnetic field due to the orbital limit, but is rather suppressed by the metamagnetic transition. Similar to the behavior at 2~K, the resistivity in the normal state is the lowest for $\mathbf{H} \parallel \mathbf{b}$. The slope of the resistivity in the normal state above $H_m$ is positive. Already for 18~deg, the resistivity is significantly reduced just above the metamagnetic transition, which is even more pronounced at 19~deg.  This defines the onset angle of the  SC3 phase  to be very close to 20~deg at 350~mK. 

Indeed, upon increasing the angle by only one degree, to 20~deg [see Fig.~\ref{f2} (middle panel)], the resistivity remains zero up to $H_{c2}=32$~T, below $H_m$, before increasing. Above $H_{o} = 36$~T, the resistivity decreases toward zero, indicating the onset of reentrant superconductivity. Zero resistivity ($\rho = 0$) is recovered above $H_r = 37.15$~T, but only within a very narrow field interval. At 21~deg, the reentrant superconductivity is much more pronounced and zero resistivity persists up to 42~T, the highest field of our measurements. A similar behavior is observed up to 31~deg.

At even larger angles [lower panel of Fig.~\ref{f2})], the $H_{c2}$ of the low field superconducting phase decreases with angle, while the field, $H_r$, at which the high-field reentrant SC3 phase emerges, increases. The maximum of the magnetoresistance $H_o$ marks the onset of the superconducting SC3 phase. For angles of 35 deg and above, this maximum increases in field, and zero resistivity is no longer accessible within our available field range up to 42~T  (see Fig.~\ref{f2}).

\begin{figure}
\includegraphics[width=\linewidth]{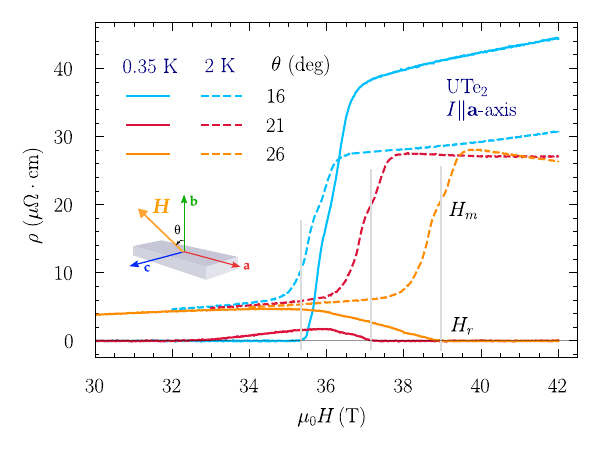}
\caption{Comparison of the magnetoresistance at $T= 2$ (dashed lines) and 0.35~K (solid lines) for three selected angles. Vertical lines indicate the field associated to the criterion $\rho =0$. For all three angles, $\rho = 0$ is within the width of the metamagnetic transition observed at $T=2$~K.}
\label{f3}
\end{figure}

In Fig.~\ref{f3} we compare the magnetoresistance at $T = 0.35$ and 2~K for 16, 21, and 26~deg. At 16 deg, the resistivity remains zero up to the the metamagnetic transition. For the other two angles, while the onset of the reentrant SC3 phase, $H_o$, occurs somewhat below the metamagnetic transition, the resistivity does not drop to zero below $H_m$. 

\begin{figure}
\includegraphics[width=\linewidth]{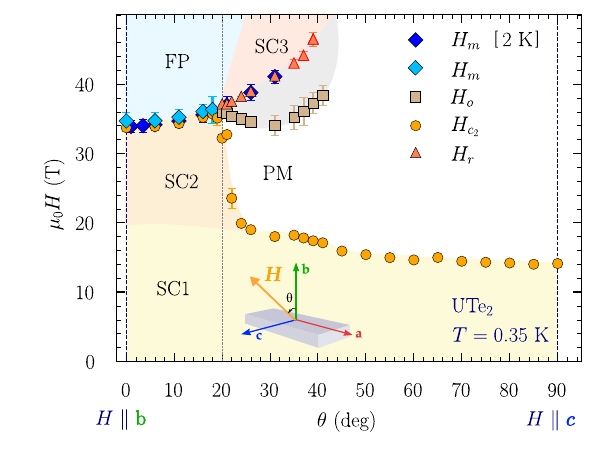}
\caption[Field-angle phase diagram of UTe2 at T=0.35 K.]{(Color online) Field-angle phase diagram of UTe$_2$ at $T=0.35~$K together with additional points of $H_m$ obtained at $2~$K. Phase transition points shown in Fig.~\ref{f1} and \ref{f2} by triangles are plotted here according to the type of the phase transition. The $H_{c2}$ is defined as the upper field-limit for which $\rho = 0$. Its error bar corresponds to the reading-uncertainty compared to the noise level.  The metamagnetic transition field  $H_m$ correspond to the middle of the magnetoresitance jump, and the error bar corresponds to the transition width. The field value of the onset of the SC3 phase $H_o$ is the local maximum of the magnetoresistance as shown in Fig.~\ref{f2}. $H_r$ is defined as the field at which the resistivity reaches zero once again. The associated error bar is obtained the same way as for $H_{c2}$. The red-edge triangles correspond to the extrapolated phase transition as shown in Fig.~\ref{f2} (lower panel).}
\label{f4}
\end{figure}

\subsection{Field-angle phase diagram}

Based on the magnetoresistance data presented above, we construct the field-angle phase diagram of UTe$_2$ in the $\bm{(bc)}$ plane shown in Fig.~\ref{f4}. The phase diagram is overall similar to previous reports \cite{Aoki2024, Wu2026}.  The metamagnetic transition field $H_m$ determined at 2~K increases with field faster than $1/\cos\theta$. At low temperature, it coincides with the upper limit of the superconducting phase for angles below 20~deg from $\bm{b}$ to the $\bm{c}$ (see the blue curve of Fig.~\ref{f3}). For angles above 20~deg, the SC3 superconducting phase occurs above $H_m$. We find that the field zero resistance $H_r$ coincides again with the metamagnetic field value $H_m$ at 2~K (see the orange and red curves of Fig.~\ref{f3}). 
Such a behavior was previously observed in other experiments not only in the $(\bm{bc})$ plane \cite{Knafo2020, Aoki2024, Thebault2026, Wu2026}, but also when there is an additional tilting towards the $\bm{a}$ axis \cite{Wu2025}. In our measurement, we are not able to determine the upper limit of the SC3 superconducting phase, except for the field applied at 21 deg.

\subsection{Temperature dependence}

\begin{figure}
\includegraphics[width=0.9\linewidth]{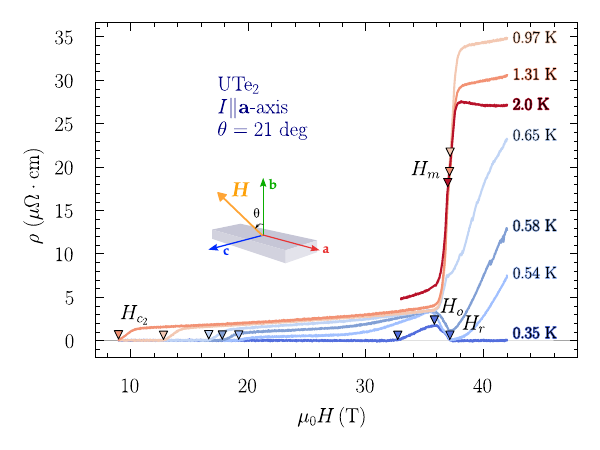}
\includegraphics[width=0.9\linewidth]{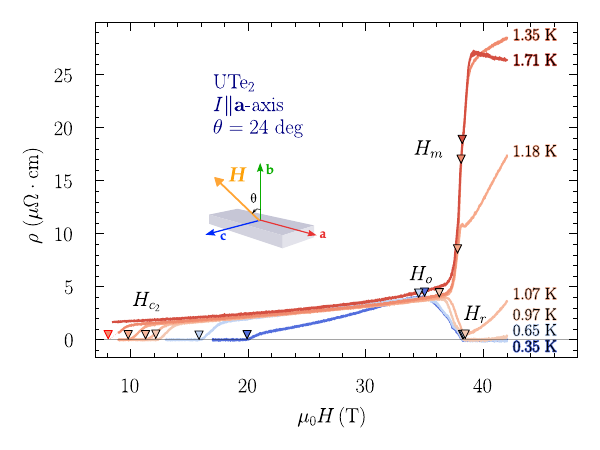}
\caption[Field-angle phase diagram of UTe2.]{
(Color online) Temperature dependence of magnetoresistance for two specific field orientations 21 deg (upper panel) and 24 deg (lower panel) between 0.35~K and $\sim 2$~K. 
Triangles mark the characteristic fields $H_{c2}$, $H_{o}$ (see Fig.~\ref{f3}), $H_r$, and $H_m$, respectively. 
}
\label{f5}
\end{figure}


Figure~\ref{f5} displays the magnetoresistance at two fixed angles for different temperatures. The upper panel shows the results obtained at  21~deg, the smallest angle at which field reentrance of superconductivity above $H_r$ occurs at $T=0.35$~K. With increasing temperature, the zero resistivity is rapidly suppressed and already at 0.58~K, the resistivity remains finite over the whole field range up to 42~T. The onset of the reentrant superconductivity, marked by the maximum in $\rho (H)$ is observed only up to 0.58~K. Already at 0.65~K the onset of the metamagnetic transition, marked by a sharp increase of the resistivity at 36~T, is observed.

The lower panel in Fig.~\ref{f5} shows the magnetoresistance at $\theta = 24$~deg.  Compared to 21~deg, the reentrant superconductivity is strongly enhanced and zero resistivity is observed all the way up to $T \approx 1$~K above $H_r$ up to the highest field of 42~T. 

\begin{figure}
\includegraphics[width=0.9\linewidth]{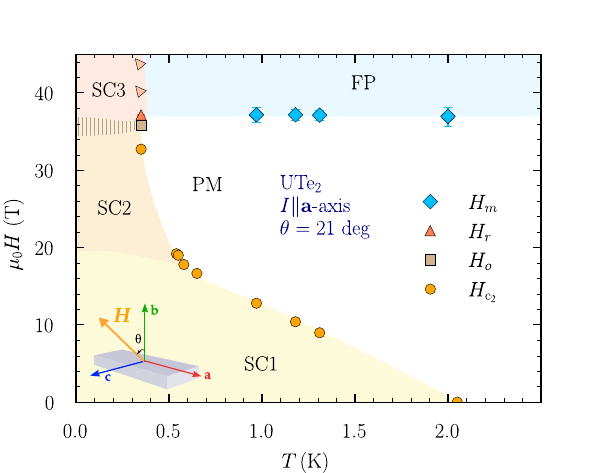}
\includegraphics[width=0.9\linewidth]{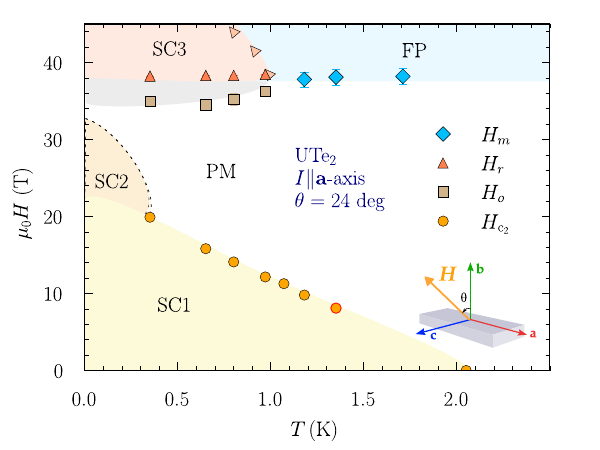}
\caption{(Color online) Field-temperature phase diagram for two specific field orientations of 21 deg (upper panel) and 24 deg (lower panel). At $\theta = 21$~deg, the field reinforced phase SC2 is clearly present, while for 24~deg it is not observed down to 0.35~K. However, our previous measurements at 24~deg at temperatures down to 50~mK~\cite{Aoki2024} suggest the presence of the SC2 phase at lower temperatures as schematically shown by a dotted line in the lower panel. }
\label{fig:f6}
\end{figure}

From the measurements of $\rho (H)$ at different temperatures, we construct the field-temperature phase diagrams shown in the upper and lower panels of Fig.~\ref{fig:f6} for 21 deg and 24~deg, respectively.

Our measurements clearly show that the field enhancement of the SC2 superconducting phase is still present at 21~deg. However, compared to $\bm{H} \parallel \bm{b}$, where the superconducting critical temperature $T_{SC2}$ of the SC2 phase has a positive slope (increasing magnetic field increases $T_{SC2}$), no field enhancement of $T_{SC2}$ is observed, and $T_{SC2}$ is much lower than $T_c$ at $H=0$~T. At $T = 0.35$~K, the resistivity is finite over a tiny field range between the suppression of the SC2 phase and the emergence of the field-induced SC3 phase in the field polarized phase (gray shaded region in the lower panel of Fig.~\ref{fig:f6}). Whether this is still the case at lower temperatures remains an important open question. The maximal critical temperature $T_{SC3}$ of the field polarized superconducting phase SC3 is similar to $T_{SC2}$, suggesting that at the critical angle of 21~deg, SC3 is the extension of SC2. 

The behavior seems to change within a small angular range. At 24~deg, the field reinforced phase SC2 is absent down to 0.35~K, and the high field SC3 phase above $H_m$ is fully separated from the low field SC1 phase. However, our previous lower temperature measurements~\cite{Aoki2024} suggest that the SC2 phase is still present up to 30~T at lower temperatures at this orientation. Compared to 21~deg, the SC3 phase is significantly enhanced with the maximum $T_{SC3}$ of about 1~K. The onset field $H_o$ of the SC3 phase increases with temperature and vanished near the maximum of $T_{SC3}$. 
\subsection{Quantum oscillations}

\begin{figure}
\includegraphics[width=0.9\linewidth]{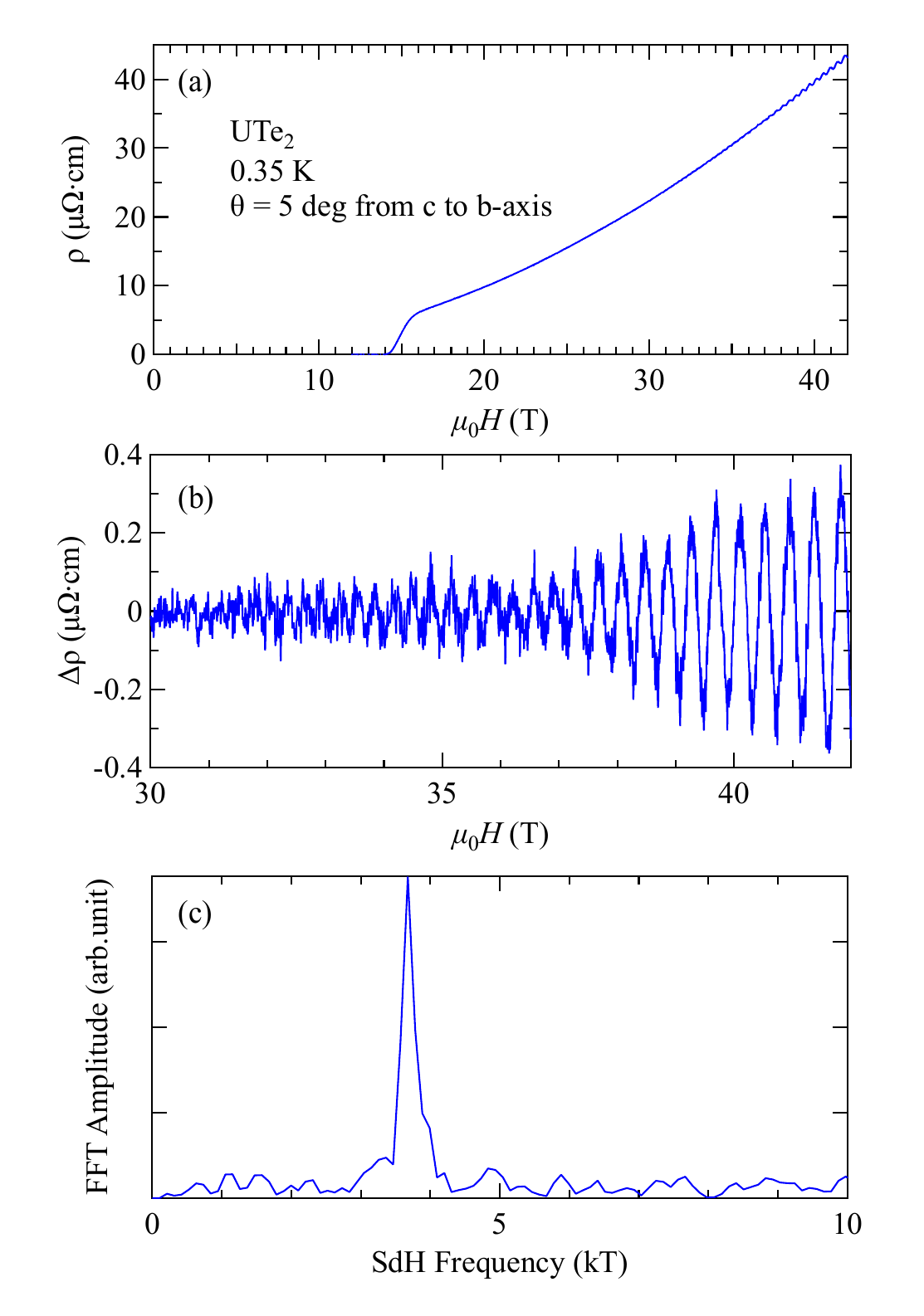}
\caption[Magnetoresistance of UTe2 at T=0.35 K.]{(Color online) (a) Magnetoresistance of UTe$_2$ at $T=0.35$~K for a field applied at 5 deg from c-axis. (b) The same signal as in (a) after subtraction of the nonoscillating background. (c) Fast Fourier Transform  spectrum of the signal shown in (b) over a field range from 30~T to 42~T.
}
\label{fig:SdH}
\end{figure}

Figure~\ref{fig:SdH}(a) shows the magnetoresistance for the field direction tilted by 5 deg from $\bm{c}$ to the $\bm{b}$-axis. At this orientation, superconductivity is suppressed at $H_{\rm c2}\sim 15\,{\rm T}$, above which the magnetoresistance increases with field, approximately as a function of $H^2$. As clearly seen in the raw data, Shubnikov-de Haas (SdH) oscillations appears above $\sim 30\,{\rm T}$. Upon subtracting the nonoscillating background, SdH oscillations become even more clearly visible, as shown in Fig.~\ref{fig:SdH}(b). The corresponding FFT spectrum is shown in Fig.~\ref{fig:SdH}(c), indicating the SdH branch $\alpha$ or $\beta$ with the frequency, $3.7\,{\rm kT}$, which approximately agrees with the previously reported values \cite{Aoki2022osc, Eaton2024}. This frequency corresponds to the cross sectional area for the main quasi-2D electron or hole Fermi surface with cylindrical shape. It should be noted that the FFT spectrum for $H\parallel c$-axis (not shown here) exhibits a splitting of the SdH frequencies, revealing branches $\alpha_1$, $\alpha_2$ and $\beta$ \cite{Aoki2022osc, Aoki2023, Aoki2024}. The corresponding warping of the Fermi surface has been confirmed recently using angle-dependent
magnetoresistance oscillations.~\cite{Kimata2026} 
Upon rotating the field angle from $\bm{c}$ to the $\bm{b}$-axis, SdH signals become smaller and smaller, and finally disappear above 20 deg.

\begin{figure}
\includegraphics[width=0.9\linewidth]{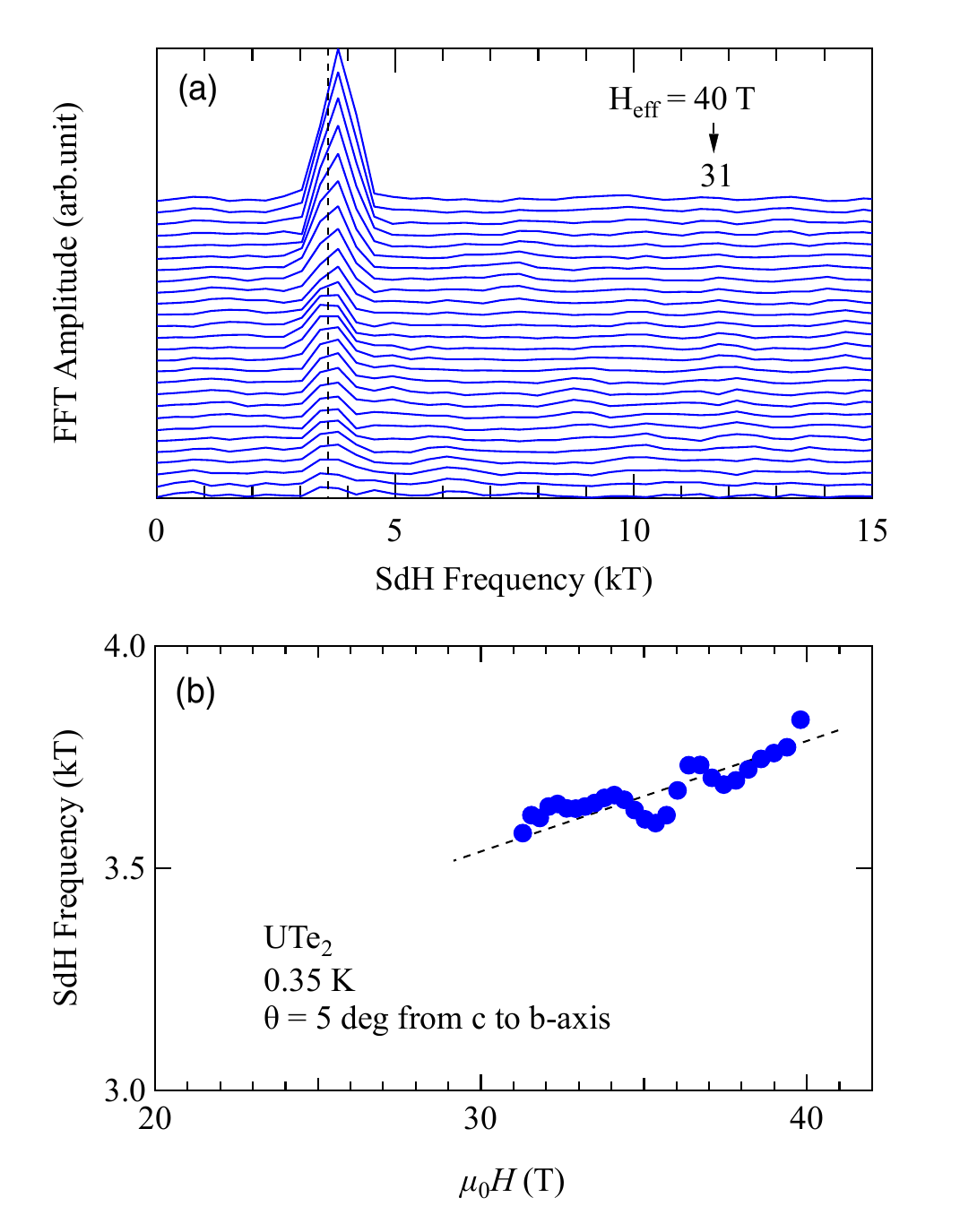}
\caption{(Color online) (a) FFT spectra of the oscillatory signal shown in Fig.~\ref{fig:SdH}(b), obtained over equal inverse-field intervals. The corresponding effective fields, $H_{eff}$, range from 31 to 40 T. (b) Field dependence of the principal SdH frequency extracted from the FFT analysis shown in (a).}
\label{fig:Hdep_F}
\end{figure}

Figure~\ref{fig:Hdep_F}(a) shows the FFT spectra obtained over different magnetic-field ranges while keeping the same $1/H$ window width. The effective field is $40\,{\rm T}$ for the upper FFT spectrum and $31\,{\rm T}$ for the lower one. The peak position shifts slightly towards higher frequencies with increasing effective field.
Figure~\ref{fig:Hdep_F}(b) displays the field dependence of the SdH frequency. As the magnetic field increases, the frequency exhibits a slight upward shift.

\section{Discussion}
The main result of this study is the detailed determination of the $H - \theta$ phase diagram shown in Fig.~\ref{f4}, and the temperature dependence of different superconducting phases at angles close to the critical angle $\theta \approx 20$~deg, where the field-enhanced SC2 superconducting phase vanishes and the field-induced phase SC3 occurs. Our results are very similar to those presented in Ref.~\citen{Wu2026}.  However,  the data shown in Ref.~\citen{Wu2026} suggest that the SC3 superconducting phase spills out beyond the first-order metamagnetic transition, i. e. the SC3 phase emerges below the transition at $H_m$ in the paramagnetic phase. On the contrary, our data rule out this possibility, as explicitly show in Fig.~\ref{f3}. Indeed, the onset of zero resistivity, taken as the strict criterion for a fully developped superconducting state, coincides with the metamagnetic transition within the experimental error bar. We do confirm, however, that the onset of the superconducting state $H_{\rm o}$ occurs already far below $H_m$. This implies that the superconducting transition to the SC3 phase is large rather than sharp. We note that  the transition from the SC3 superconducting state to the normal state at its $H_{c2}$ was also reported to be extremely broad in both transport measurements \cite{Ran2019, Knafo2020, Frank2024, Helm2024, Wu2025, Thebault2026} and bulk experiments \cite{Marquardt2025}.

The origin of the very broad superconducting transition to the SC3 phase is still an open question. It is attractive to suggest that magnetic fluctuations act as a pairing glue for the SC3 phase. Enhancement of the spin fluctuations on approaching the metamagnetic transition has been studied in detail for $\bm{H} \parallel \bm{b}$. Direct evidence for strongly enhanced longitudinal electronic spin fluctuations comes from high field NMR relaxation experiments\cite{Tokunaga2022}. Furthermore, this increase of the fluctuations gives rise to a strong increase of both the Fermi-liquid $A$ coefficient of the resistivity \cite{Knafo2019, Knafo2020} and the specific heat coefficient $\gamma$ \cite{Imajo2019, AMiyake2021JPSJ, Rosuel2023} on approaching $H_m$. Above $H_m$, both $A$ and $\gamma$ coefficients decrease abruptly. In contrast, the behavior is markedly different over the angular range from 23 deg to 40 deg in the $\bm{bc}$ plane, where the field-induced superconducting phase SC3 emerges above the metamagnetic transition. At these angles, both coefficients increase step-like at $H_m$, resulting in a regime of enhanced fluctuations above the transition \cite{Knafo2020, AMiyake2021JPSJ, Thebault2026}. This distinct field dependence of the fluctuations might explain why superconductivity can reappear above the metamagnetic transition in the $\bm{bc}$ plane. The orbital limit of superconductivity is highly sensitive to these changes, as $H_{c2}(T)|_{T \to 0}$ scales with the field-dependent effective mass $m^*(H)$ and the associated transition temperature $T_c(m^*)$. A very similar mechanism has been already discussed for the field-reentrance of superconductivity in URhGe and UCoGe where a field-dependent pairing strength is taken into account.\cite{Miyake2008c, Wu2017}

In difference to this, recent transport and Hall effect investigations\cite{Helm2024} proposed a Jaccarino-Peter-type mechanism suggesting that a field-induced internal exchange field compensates for the external field within the polarized state. Within this hypothesis, the field-polarized state above $H_m$ was argued to generate a static internal negative exchange field ($H_{\text{ex}}$) via the $5f$ uranium moments cancelling the applied field. Then, the vector sum of the applied field and the internal exchange field drops dramatically above $H_m$, creating a low effective field environment that allows for the emergence of SC3. 

Surprisingly, we identify a critical angle near 20~deg, where the SC2 phase below $H_m$ is suppressed and the SC3 phase develops at higher angles. A detailed microscopic picture for this is still missing. To discern different mechanisms for the appearance of SC3, detailed fine-tuned angle-dependent thermodynamic and, especially, microscopic experiments are still missing.

The observed SdH oscillations close to the $\bm{c}$ axis are in excellent agreement with our previous studies \cite{Aoki2022osc, Aoki2023, Aoki2024}. The extended magnetic-field range, however, allows a more detailed investigation of their field dependence. While SdH and dHvA frequencies are generally expected to be field independent, the true frequencies, $F_{\rm tr}$, of the spin-up and spin-down bands split under an applied magnetic field owing to the Zeeman effect. The experimentally observed frequency, $F_{\rm obs}$, is obtained by back-projecting $F_{\rm tr}(H)$ to zero field. As a result, $F_{\rm obs}$ remains essentially field independent as long as $F_{\rm tr}$ varies linearly with magnetic field.

The result shown in Fig.~\ref{fig:Hdep_F}(b) suggests that the true frequency exhibits a nonlinear field dependence. For $H \parallel c$, the magnetization increases nearly linearly up to 55\,{\rm T}, with only a slight convex curvature at high fields. Together, these observations may indicate field-induced modifications of the Fermi surface in UTe$_2$.

The effective mass associated with this branch is estimated to be $30$--$40\,m_0$. In general, heavy-mass Fermi-surface sheets are particularly susceptible to magnetic-field-induced modifications. Therefore, although the overall topology is expected to remain similar to that at low fields, our results suggest the presence of minor field-induced modifications of the Fermi surface in UTe$_2$.

An open question is the topology of the Fermi surface above the metamagnetic transition field $H_m$. In many heavy-fermion systems, a field-induced metamagnetic transition is accompanied by a drastic reconstruction of the Fermi surface~\cite{HAoki2014,McCollam2021,Terashima1997}. In UTe$_2$, however, the Fermi surface in the polarized phase above $H_m$ remains largely unknown. Evidence for a Fermi-surface reconstruction has been reported from Hall-effect and thermoelectric-power measurements, which reveal a significant change in carrier concentration~\cite{Niu2020}. The possible role of such a Fermi-surface reconstruction in the emergence of the SC3 phase remains to be clarified and should be investigated in future studies.

\section{Conclusions and Outlook}

In summary, we performed transport measurements on a high-quality single crystal of UTe$_2$ in static magnetic fields up to 42~T and at temperatures down to 0.35~K. We confirm that the field-induced SC3 phase emerges above a critical angle of about 20 deg, corresponding to the maximum angle at which the field-reinforced superconducting phase SC2 persists up to, or just below, the metamagnetic transition at $H_m$. Importantly, we unambiguously demonstrate that the full establishment of the SC3 phase, defined by the onset of zero resistivity, occurs only above the metamagnetic transition, although its onset is observed slightly below $H_m$. This finding indicates that the emergence of the SC3 phase is closely linked to the metamagnetic transition. 

Based solely on our magnetoresistance measurements, we cannot discriminate between the proposed mechanisms for the SC3 phase, including quantum-critical magnetic fluctuations and a field-compensation mechanism. Additional thermodynamic and microscopic studies are required to provide key evidence and elucidate the pairing mechanism responsible for the SC3 phase. The extension of the accessible magnetic-field range to 42 T and beyond will enable future experiments aimed at probing the properties of this high-field superconducting state.

Finally, quantum oscillations observed for magnetic fields close to the $\bm{c}$ axis yield a frequency of approximately 3.7~kT, consistent with previous reports. The weak field dependence of this frequency suggests moderate field-induced modifications of the Fermi surface in UTe$_2$.

\section*{Acknowledgements}
We thank W.~Knafo, D. Braithwaite, and J.-P.~Brison for fruitful discussions and critical reading. 
We acknowledge financial support from the French National Agency for Research ANR within the projects SCATE No. ANR-22-CE30-0040 and KONWEY (ANR-24-CE97-0006), from the CEA Exploratory program 23P61 ToSuFe, and from the JSPS programs KAKENHI (JP19H00646, JP20K20889, JP20H00130, JP20KK0061, JP20K03854, JP22H04933). We acknowledge support of the LNCMI-CNRS, member of the European Magnetic Field Laboratory (EMFL), and from the Laboratoire d’excellence LANEF (ANR-10-LABX-0051).



%


\bibliographystyle{jpsj}
\bibliography{UTe2_biblio}

\end{document}